\documentclass[12pt,fleqn]{article}
\usepackage{graphicx,cite,amssymb}

\newcommand{\newsection}[1]{\section{#1}\setcounter{equation}{0}}

\def\be{\begin{equation}}
\def\ee{\end{equation}}
\def\bea{\begin{eqnarray}}
\def\eea{\end{eqnarray}}
\def\nnb{\nonumber}
\def\bbuildrel#1_#2^#3{\mathrel{\mathop{\kern 0pt#1}\limits_{#2}^{#3}}}
\def\slash#1{\setbox0=\hbox{$#1$}#1\hskip-\wd0\dimen0=5pt\advance
       \dimen0 by-\ht0\advance\dimen0 by\dp0\lower0.5\dimen0\hbox
         to\wd0{\hss\sl/\/\hss}}

\newcommand{\f}{\frac}
\newcommand{\fm}[2]{{\textstyle \frac{#1}{#2}}}
\newcommand{\me}[1]{\langle#1\rangle}
\newcommand{\al}{\alpha_{\mathrm s}}
\newcommand{\alt}{\widetilde{\alpha}_{\mathrm s}}
\newcommand{\ep}{\epsilon}

\begin{document}

\begin{titlepage}

\begin{flushright}
TTP17-007\\
IFT-2/2017\\[2cm]
\end{flushright}
\begin{center}
\setlength {\baselineskip}{0.3in} 
{\bf\Large\boldmath NNLO QCD counterterm contributions to $\bar B \to X_s \gamma$
                    for the physical value of $m_c$}\\[15mm]
\setlength {\baselineskip}{0.2in}
{\large  Miko{\l}aj Misiak$^{1,2}$, Abdur Rehman$^{1,3}$ and~ Matthias Steinhauser$^4$}\\[5mm]
$^1$~{\it Institute of Theoretical Physics, Faculty of Physics, University of Warsaw,\\
                    02-093 Warsaw, Poland.}\\[5mm]
$^2$~{\it Theoretical Physics Department, CERN, CH-1211 Geneva 23, Switzerland.}\\[5mm]
$^3$~{\it National Centre for Physics, Quaid-i-Azam University Campus,\\
              Islamabad 45320, Pakistan.}\\[5mm]
$^4$~{\it Institut f\"ur Theoretische Teilchenphysik, 
          Karlsruhe Institute of Technology (KIT),\\
          76128 Karlsruhe, Germany.}\\[3cm] 
{\bf Abstract}\\[5mm]
\end{center} 
\setlength{\baselineskip}{0.2in} 

One of the most important ${\mathcal O}(\al^2)$ corrections to the $\bar B\to
X_s\gamma$ branching ratio originates from interference of contributions from
the current-current and photonic dipole operators. Its value has been
estimated using an interpolation in the charm quark mass between the known
results at $m_c=0$ and for $m_c \gg \f12 m_b$. An explicit calculation for the
physical value of $m_c$ is necessary to remove the associated uncertainty.  In
the present work, we evaluate all the ultraviolet counterterm contributions
that are relevant for this purpose.

\end{titlepage}

\newsection{Introduction \label{sec:intro}}

The inclusive decay $\bar B\to X_s\gamma$ belongs to the most interesting
flavour changing processes.  Many popular extensions of the Standard Model
(SM) receive important constraints from its branching ratio ${\mathcal
B}_{s\gamma}$.  For instance, the 95\%$\,$C.L.\ bound it provides on the
charged Higgs boson mass in the Two-Higgs-Doublet Model II is in the vicinity
of $580\,$GeV~\cite{Belle:2016ufb,Misiak:2017bgg}. Strengthening such bounds
in the future is likely to be achieved with the upcoming precise measurements
of ${\mathcal B}_{s\gamma}$ at Belle~II~\cite{Aushev:2010bq}, as well as with
more accurate theoretical calculations within the SM.

The SM prediction for the CP- and isospin-averaged branching ratio ${\mathcal
B}_{s\gamma}$ amounts to~\cite{Misiak:2015xwa,Czakon:2015exa}
\be \label{brsm}
{\mathcal B}_{s\gamma}^{\rm SM} = (3.36 \pm 0.23) \times 10^{-4},
\ee
for $E_\gamma > E_0 = 1.6\,$GeV. The corresponding experimental world average
evaluated by HFAG~\cite{Amhis:2016xyh} reads
\be \label{brHFAG}
{\mathcal B}_{s\gamma}^{\rm exp} = (3.32 \pm 0.15) \times 10^{-4}.
\ee
It is based on the Belle~\cite{Belle:2016ufb,Saito:2014das},
Babar~\cite{Aubert:2007my,Lees:2012ym,Lees:2012wg} and
CLEO~\cite{Chen:2001fja} measurements performed at $E_0 \in [1.7,2.0]\,$GeV,
and then extrapolated down to $E_0 = 1.6\,$GeV using the method of
Ref.~\cite{Buchmuller:2005zv}.  A recent discussion of the energy
extrapolation issue and its effect on the averages can be found in
Ref.~\cite{Misiak:2017bgg}. In that paper
\be \label{br19}
{\mathcal B}_{s\gamma}^{\rm exp} = (3.27 \pm 0.14) \times 10^{-4}
\ee
was obtained by naively averaging the measurements at $E_0 = 1.9\,$GeV 
only~\cite{Belle:2016ufb,Saito:2014das,Aubert:2007my,Lees:2012ym,Lees:2012wg},
and then performing an extrapolation to $E_0 = 1.6\,$GeV.

As far as the SM prediction is concerned, it is based on the fact that the
hadronic decay rate $\Gamma(\bar B \to X_s \gamma)$ is well approximated by
the corresponding partonic one $\Gamma(b \to X_s^p \gamma)$, where $X_s^p = s,
sg, sgg, sq\bar q,\ldots$. The difference between the two rates is due to
non-perturbative effects. Such effects contribute at around $+3\%$
level~\cite{Buchalla:1997ky} to the central value in Eq.~(\ref{brsm}), while
the associated uncertainty is estimated~\cite{Benzke:2010js} to be around $\pm
5\%$. This uncertainty is combined in quadrature with the remaining three ones
that are due to the input parameters ($\pm 2\%$), unknown higher-order
${\mathcal O}(\al^3)$ perturbative effects ($\pm 3\%$), as well an
interpolation in the charm quark mass $m_c$ ($\pm 3\%$). The interpolation is
used to estimate some of the most important ${\mathcal O}(\al^2)$ corrections.

The purpose of our present work is providing a contribution to removing the
latter uncertainty. Perturbative corrections to $\Gamma(b \to X_s^p \gamma)$
are most conveniently analyzed in the framework of an effective theory that
arises after decoupling the $W$ boson and all the heavier particles. The
relevant weak interaction Lagrangian is then given by a linear combination of
four-quark and dipole-type operators:~
${\mathcal L}_{\rm weak} \sim \sum_i C_i (\mu) Q_i$
(see, e.g., Eq.~(1.6) of Ref.~\cite{Czakon:2015exa}). For most of our
discussion here, only three of them are going to be relevant, namely
\mathindent0cm
\be
Q_1  = (\bar{s}_L \gamma_{\mu} T^a c_L) (\bar{c}_L     \gamma^{\mu} T^a b_L),\hspace{8mm}
Q_2  = (\bar{s}_L \gamma_{\mu}     c_L) (\bar{c}_L     \gamma^{\mu}     b_L),\hspace{8mm}
Q_7  =  \f{em_b}{16\pi^2} (\bar{s}_L \sigma^{\mu \nu}     b_R) F_{\mu \nu}.
\ee
\mathindent1cm
The $\overline{\rm MS}$-renormalized Wilson coefficients $C_i(\mu)$ are first
evaluated at the electroweak scale $\mu_0 \sim M_W,m_t$, and then evolved down
to $\mu_b \sim m_b$ using the effective theory renormalization group
equations. A complete ${\mathcal O}(\al^2)$ calculation of all the relevant
$C_i(\mu_b)$ was finalized a decade ago, with the latest contribution
amounting to a determination of the necessary four-loop anomalous dimension
matrix~\cite{Czakon:2006ss}.

Once the Wilson coefficients $C_i(\mu_b)$ are found, one writes the
perturbative rate as\footnote{
For simplicity, we ignore here the higher-order electroweak and ${\mathcal
O}(V_{ub})$ corrections. They are included in the numerical evaluation of the
SM prediction~(\ref{brsm}). Moreover, we make use of the fact that 
$C_i \in \mathbb{R}$ in the SM.}
\be \label{rate}
\Gamma(b \to X_s^p \gamma)_{E_\gamma > E_0} ~=~ 
\f{G_F^2 m_b^5 \alpha_{\mathrm em}}{32 \pi^4} \left|V_{ts}^* V_{tb} \right|^2
\sum_{i,j=1}^8 C_i(\mu_b) \, C_j(\mu_b) \, \hat{G}_{ij},
\ee
where $G_F$ is the Fermi constant, and $V_{ij}$ are the
Cabibbo-Kobayashi-Maskawa matrix elements.  Both the $b$-quark mass $m_b$ and
the electromagnetic coupling $\alpha_{\mathrm em}$ are assumed to be on-shell
renormalized. 

\begin{figure}[t]
\begin{center}
\includegraphics[width=59mm,angle=0]{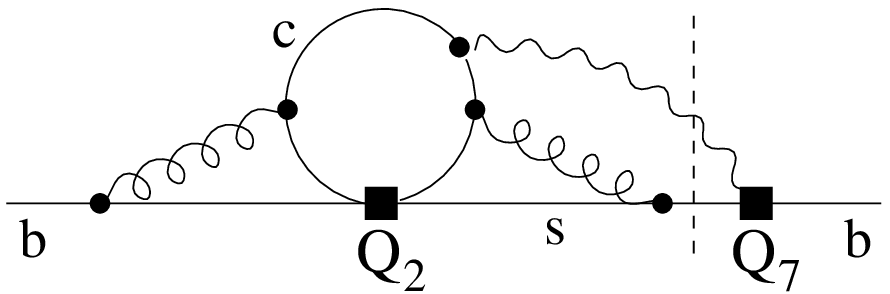}
\includegraphics[width=58mm,angle=0]{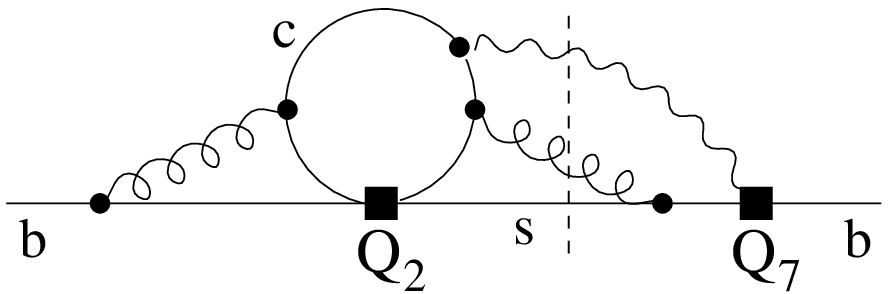}
\includegraphics[width=44mm,angle=0]{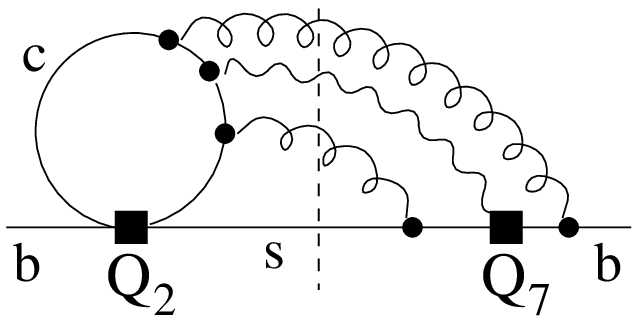}
\caption{\sf Sample diagrams for $\hat{G}^{(2)\rm bare}_{27}$
with unitarity cuts indicated by the dashed lines. \label{fig:NNLO}}
\end{center}
\end{figure}

The interference terms $\hat{G}_{ij}$ form a symmetric $8 \times 8$
matrix. Their perturbative expansion in $\alt \equiv \al(\mu_b)/(4\pi)$ reads
\be
\hat{G}_{ij} = \hat{G}^{(0)}_{ij} + \alt\, \hat{G}^{(1)}_{ij} + \alt^2\, \hat{G}^{(2)}_{ij} + {\mathcal O}(\alt^3).
\ee
The Next-to-Next-to-Leading Order (NNLO) corrections to be 
studied in the present work are given by $\hat{G}^{(2)}_{17}$ and
$\hat{G}^{(2)}_{27}$.  They describe interferences of decay amplitudes
generated by the current-current operators $Q_{1,2}$ and the photonic dipole
operator $Q_7$. Such interference terms are most conveniently represented via
Cutkosky rules as four-loop propagator diagrams with unitarity cuts. Sample
diagrams are presented in Fig.~\ref{fig:NNLO}. Altogether, around 850 of such
four-loop diagrams need to be evaluated.

Despite the fact that $\hat{G}_{(1,2)7}$ determine some of the most important
NNLO corrections to ${\mathcal B}_{s\gamma}$ (comparable to the present
experimental and theoretical uncertainties), their explicit form for the
physical value of $m_c$ remains unknown, which is due to extreme complexity of
the necessary calculation.  What has been found so far for arbitrary
$m_c$ are contributions from diagrams with massless and massive quark loops on
the gluon lines~\cite{Ligeti:1999ea,Bieri:2003ue,Boughezal:2007ny}. They are
the basis for evaluating $\hat{G}_{(1,2)7}$ in the Brodsky-Lepage-Mackenzie
(BLM)~\cite{Brodsky:1982gc} approximation. As far as the non-BLM terms are
concerned, they have been calculated for $m_c \gg \f12
m_b$~\cite{Misiak:2010sk} and at $m_c=0$~\cite{Czakon:2015exa} only. Next,
their values for the physical $m_c$ were found using an interpolation, as
described in detail in Ref.~\cite{Czakon:2015exa}.

The interpolated contributions affect ${\mathcal B}_{s\gamma}$ by $(+5 \pm
3)\%$, which implies that their effects on both the central value and the
overall uncertainty are sizeable. The interpolation uncertainty could be
completely removed via a calculation of $\hat{G}_{(1,2)7}$ directly at the
physical value of $m_c$. Such a calculation involves the ``bare'' diagrams
like the ones in Fig.~\ref{fig:NNLO}, as well as many lower-loop diagrams with
insertions of ultraviolet counterterms.

In the present paper, we provide results for all the necessary counterterm
contributions in the $m_c\neq 0$ case. Although it is only a first step
towards the complete NNLO calculation of $\hat{G}_{(1,2)7}$, it is by no means
a trivial one given that two mass scales are present in the relevant
three-loop propagator diagrams with unitarity cuts.\footnote{
The two mass scales are given by $m_b$ and $m_c$, while the s-quark is treated
as massless throughout the paper.}
After generating the considered interference contributions and expressing them
in terms of Master Integrals (MIs), we evaluate all the MIs using several
methods. In the first approach, we derive a system of Differential Equations (DEs)
for the MIs, and solve it numerically starting from boundary conditions at
$m_c \gg \f12 m_b$. The latter are found using asymptotic expansions in such a
limit. The relevance of this method lies in the fact that it is basically the
only feasible one for the future ``bare'' calculation. Thus, our current task
serves as a preparation for the future enterprise. Apart from the numerical
approach, we also use analytical methods for determining all the necessary
MIs, either in closed forms or as expansions in $z = m_c^2/m_b^2$.

Similarly to the $m_c=0$ calculation of Ref.~\cite{Czakon:2015exa}, we
restrict ourselves to the limit $E_0=0$. Considering such a limit is a
necessary step on the way to the $E_0 \neq 0$ case. Results in the latter case
are going to be obtained in the future by subtracting the low-$E_\gamma$ tail
from the total rate, in the same way as it was done for
$\hat{G}_{77}$~\cite{Blokland:2005uk,Melnikov:2005bx}.

Our article is organized as follows. In Section~\ref{sec:renor}, we define all
the relevant counterterm contributions via an explicit formula for the
renormalized $\hat{G}_{(1,2)7}$. Expressions for them in terms of the MIs, as
well as DEs for the MIs are given in Section~\ref{sec:de}.  Our numerical
results for all the considered counterterm contributions are presented in
Section~\ref{sec:num}. Section~\ref{sec:fin} is devoted to presenting our
final results as expansions in $z$. We conclude in Section~\ref{sec:sum}.

\newsection{\boldmath The NNLO renormalization formula for $\hat{G}_{17}$ and $\hat{G}_{27}$ \label{sec:renor}}

The renormalized $\hat{G}^{(2)}_{i7}$ for $i=1,2$ are obtained from the bare ones
$\hat{G}_{i7}^{(2)\rm bare}$ and the ultraviolet counterterms. The necessary
expression in the $m_c=0$ case together with all its ingredients was presented
in Section~2.2 of Ref.~\cite{Czakon:2015exa}.  Following that article, we skip
the contributions that are due to charm quark loops on the gluon lines
together with the corresponding counterterms. Such NNLO contributions have
been already calculated for arbitrary $m_c$ in Ref.~\cite{Boughezal:2007ny},
and are included in the phenomenological analysis of
Ref.~\cite{Czakon:2015exa} (Eq.~(3.8) there). Under such assumptions, the
renormalization formula for arbitrary $m_c$ takes the following form (for $i=1,2$):
\bea
\alt\, \hat{G}_{i7}^{(1)} + \alt^2\, \hat{G}_{i7}^{(2)} &=& 
Z_b^{\rm OS}\, Z_m^{\rm OS}\, Z_{77} \bigg\{ \alt^2\, s^{3\ep}\, \hat{G}_{i7}^{(2) \rm bare} 
+ (Z_m^{\rm OS}-1)\, s^\ep \left[ Z_{i4}\, \hat{G}_{47}^{(0)m}  
+ \alt\, s^\ep\, \hat{G}_{i7}^{(1)m} \right] \nnb\\[2mm] 
&+& \alt\, (Z_G^{\rm OS}-1)\, s^{2\ep}\, \hat{G}_{i7}^{(1)3P} +
Z_{i7}\, Z_m^{\rm OS} \left[ \hat{G}_{77}^{(0)} + \alt\, s^{\ep}\, \hat{G}_{77}^{(1)\rm bare} \right]\nnb\\[2mm] 
&+& \alt\, Z_{i8}\, s^\ep\, \hat{G}_{78}^{(1)\rm bare} 
+ \sum_{j=1,\ldots,6,11,12} Z_{ij}\, s^\ep \left[ \hat{G}_{j7}^{(0)} 
+ \alt\, s^\ep\, Z_g^2\, \hat{G}_{j7}^{(1)\rm bare} \right]\nnb\\[2mm] 
&+&  2\alt s^{2\ep} (Z_m-1)\, z \f{d}{dz}\, \hat{G}_{i7}^{(1)\rm bare} \bigg\} +~ {\mathcal O}(\alt^3),
\label{renor}
\eea
It turns out to differ from the $m_c=0$ case only in the last line where the
renormalization of $m_c$ is taken into account. Moreover, we have simplified a
few details of the notation without changing the actual content of the
equation. Its l.h.s.\ corresponds to such a renormalization scheme where the
Wilson coefficients, $\alpha_s$ and $m_c$ are $\overline{\rm MS}$-renormalized
at the scale $\mu_b$. On the other hand, the $b$-quark mass and the external
quark fields are on-shell renormalized, which is indicated by the superscript
${\rm OS}$ at the corresponding renormalization constants. In these constants
(Eq.~(2.8) of Ref.~\cite{Czakon:2015exa}), as well as in Eq.~(\ref{renor})
here, one should substitute $s=\mu_b^2/m_b^2$. The remaining renormalization
constants are the ${\rm MS}$-scheme ones (Eq.~(2.9) of
Ref.~\cite{Czakon:2015exa}).  In particular $Z_m = 1 -4\alt/\ep + {\mathcal
O}(\alt^2)$~ in~ $D = 4-2\ep$~ spacetime dimensions.

The bare $\hat{G}_{kl}$ on the r.h.s.\ of Eq.~(\ref{renor}) are assumed to
have been evaluated for the renormalization scale $\mu^2 = e^\gamma
m_b^2/(4\pi)$, where $\gamma$ is the Euler-Mascheroni constant. They involve
indices corresponding to the operators listed in Eqs.~(1.6) and (2.5) of
Ref.~\cite{Czakon:2015exa}. In some of the cases, the superscript ``${\rm
bare}$'' has been replaced by either ``$3P$'' or ``$m$''. In these cases,
either only the three-particle cuts were included ($3P$), or one of the
$b$-quark propagators was squared ($m$). The latter operation is the simplest
way to obtain the counterterms that originate from renormalization of the
$b$-quark mass.
\begin{figure}[t]
\begin{center}
\includegraphics[width=45mm,angle=0]{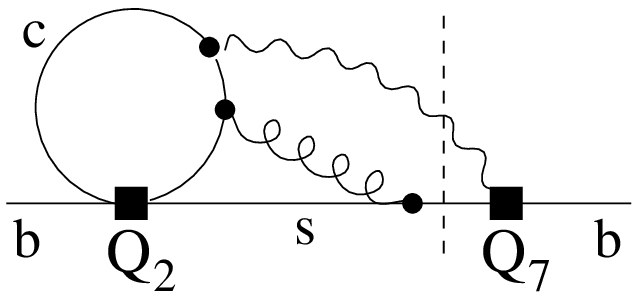}\hspace{1cm}
\includegraphics[width=45mm,angle=0]{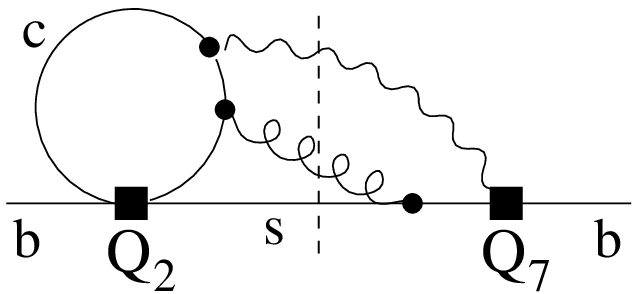}
\caption{\sf Sample diagrams for $\hat{G}^{(1)\rm bare}_{27}$
with unitarity cuts indicated by the dashed lines. \label{fig:NLO}}
\end{center}
\end{figure}

Explicit expressions for all the necessary bare $\hat{G}_{kl}$ in the $m_c=0$
case have been given in Eqs.~(2.3)-(2.7) of Ref.~\cite{Czakon:2015exa}. Only a
few of them get modified in the $m_c\neq 0$ case, namely
\bea
&& \hat{G}_{i7}^{(2) \rm bare},  \hat{G}_{i7}^{(1) \rm bare},  
\hat{G}_{i7}^{(1)3P},  \hat{G}_{i7}^{(1)m}, \mbox{~~for~~} i=1,2,\nnb\\
&& \hat{G}_{7j}^{(1) \rm bare}, \mbox{~~for~~} j=11,12.
\eea
In the following, we shall calculate all of them except for
$\hat{G}_{17}^{(2)\rm bare}$ and $\hat{G}_{27}^{(2)\rm bare}$.  Some of the
considered quantities are related to each other in a very simple manner. In
particular
\mathindent0cm
\begin{displaymath}
\hat{G}_{17}^{(1)\rm bare}  = -\f16 \hat{G}_{27}^{(1)\rm bare}, \hspace{8mm}  
\hat{G}_{17}^{(1)3P}  = -\f16 \hat{G}_{27}^{(1)3P}, \hspace{8mm}  
\hat{G}_{17}^{(1)m}  = -\f16 \hat{G}_{27}^{(1)m}, \hspace{8mm}  
\hat{G}_{7(11)}^{(1)\rm bare}  = -\f16 \hat{G}_{7(12)}^{(1)\rm bare}.
\end{displaymath}
\mathindent1cm
It is due to the fact that in all the non-vanishing Feynman diagrams that
contribute to these quantities (e.g., those in Fig.~\ref{fig:NLO}), the internal gluon
couples only once to the charm quark loop. In consequence, the diagrams
with $Q_1$ and $Q_2$ differ only by a colour factor of $-\f16$ that comes from
the identity $T^a T^b T^a = -\f16 T^b$ for the $SU(3)$ generators. The case is
analogous for the evanescent operators
\bea
Q_{11} &=& (\bar{s}_L \gamma_{\mu_1}
                      \gamma_{\mu_2}
                      \gamma_{\mu_3} T^a c_L)(\bar{c}_L \gamma^{\mu_1}
                                                        \gamma^{\mu_2}
                                                        \gamma^{\mu_3} T^a b_L)
-16 Q_1,\nnb\\[1mm]
Q_{12} &=& (\bar{s}_L \gamma_{\mu_1}
                      \gamma_{\mu_2}
                      \gamma_{\mu_3}     c_L)(\bar{c}_L \gamma^{\mu_1}
                                                        \gamma^{\mu_2}
                                                        \gamma^{\mu_3}     b_L)
-16 Q_2.
\eea

Another simple identity allows us to express $\hat{G}_{7(12)}^{(1)\rm bare}$ in
terms of various contributions to $\hat{G}_{27}^{(1)\rm bare}$. It reads
\be \label{g712}
\hat{G}_{7(12)}^{(1)\rm bare} = -4\ep \left\{ (1+\ep) \hat{G}_{27}^{(1)2P(d)}
+ (5+\ep) \left[ \hat{G}_{27}^{(1)2P(u)} + \hat{G}_{27}^{(1)3P} \right] \right\},
\ee
where $\hat{G}_{27}^{(1)2P(d)}$ and $\hat{G}_{27}^{(1)2P(u)}$ stand for the
two-particle cut contributions to $\hat{G}_{27}^{(1)\rm bare}$ that are
proportional to the down-type and up-type quark electric charges,
respectively. The identity (\ref{g712}) can be recovered after expressing
all the involved quantities in terms of the MIs, but without actually
calculating any of the MIs.

Thus, what remains to be calculated are the six quantities on the r.h.s.\ of
the following two equations: 
\bea 
\hat{G}_{27}^{(1)\rm bare} &=& \hat{G}_{27}^{(1)2P(d)} + \hat{G}_{27}^{(1)2P(u)} + \hat{G}_{27}^{(1)3P},\label{g27}\\
\hat{G}_{27}^{(1)m} &=& \hat{G}_{27}^{(1)m,2P(d)} + \hat{G}_{27}^{(1)m,2P(u)} + \hat{G}_{27}^{(1)m,3P}.\label{g27m}
\eea
Analogously to the first of the above equations, the three terms on the
r.h.s.\ of the second one are defined according to the quark electric charges,
and to the number of final-state particles.

All the considered quantities need to be found up to ${\mathcal O}(\ep)$. In
the case of Eq.~(\ref{g27}), it is equivalent to extending the NLO
calculations of
Refs.~\cite{Ali:1990tj,Greub:1996jd,Greub:1996tg,Buras:2001mq,Buras:2002tp}
to one more order in $\ep$.  Such an extension is actually already available
from Ref.~\cite{Asatrian:2005pm}, and we shall compare our results to that
article. As far as the counterterms in Eq.~(\ref{g27m}) for $m_c\neq 0$ are
concerned, our calculation is new.
\begin{table}[t]
\begin{center}
\begin{tabular}{|clcr|clcr|clcr|}\hline
&&&&&&&&&&&\\
$I_1$ &&&& $I_7$ &&&& $I_{13}$ &&& \\[-8mm]
&& \includegraphics[width=2cm,angle=0]{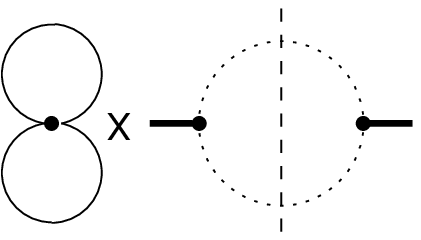} &&
&& \includegraphics[width=2cm,angle=0]{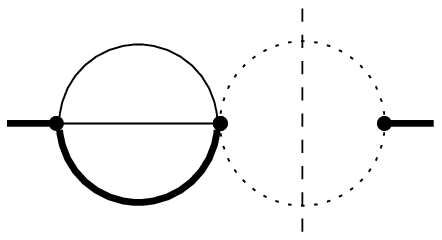} &&
&& \includegraphics[width=2cm,angle=0]{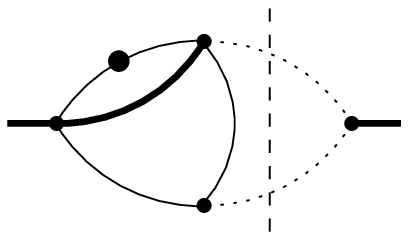} & \\[4mm]
$I_2$ &&&& $I_8$ &&&& $I_{14}$ &&& \\[-8mm]
&& \includegraphics[width=2cm,angle=0]{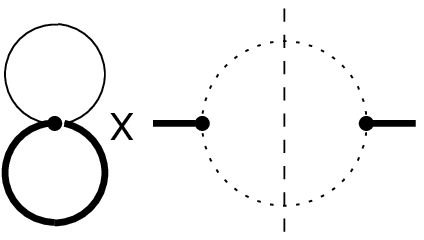} &&
&& \includegraphics[width=2cm,angle=0]{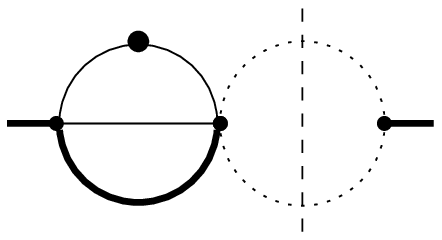} &&
&& \includegraphics[width=2cm,angle=0]{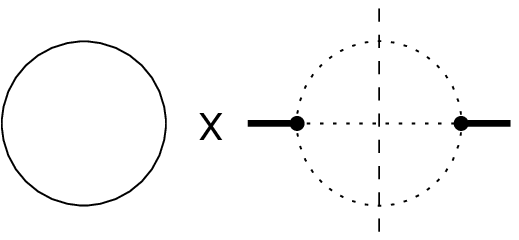} & \\[4mm]
$I_3$ &&&& $I_9$ &&&& $I_{15}$ &&& \\[-8mm]
&& \includegraphics[width=2cm,angle=0]{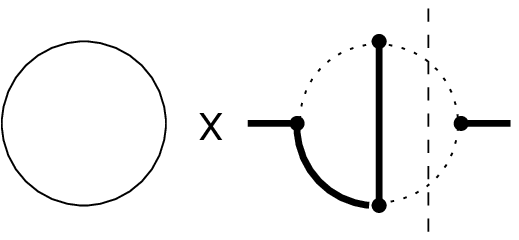} &&
&& \includegraphics[width=2cm,angle=0]{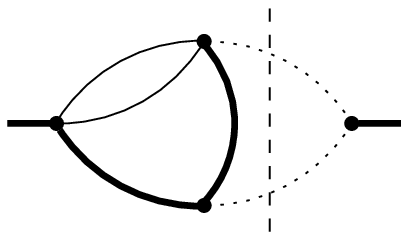} &&
&& \includegraphics[width=2cm,angle=0]{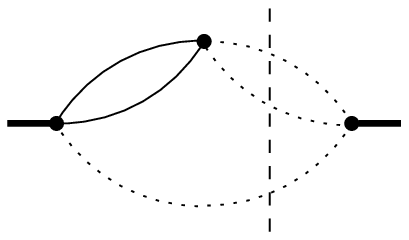} & \\[4mm]
$I_4$ &&&& $I_{10}$ &&&& $I_{16}$ &&& \\[-8mm]
&& \includegraphics[width=2cm,angle=0]{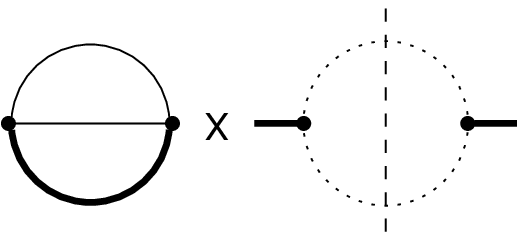} &&
&& \includegraphics[width=2cm,angle=0]{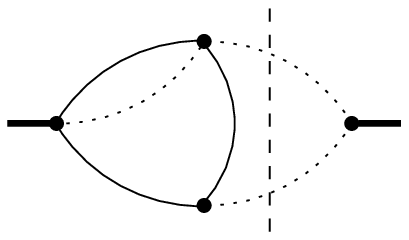} &&
&& \includegraphics[width=2cm,angle=0]{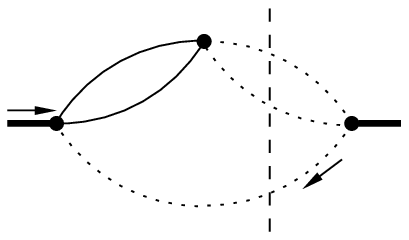} & \\[4mm]
$I_5$ &&&& $I_{11}$ &&&& $I_{17}$ &&& \\[-8mm]
&& \includegraphics[width=2cm,angle=0]{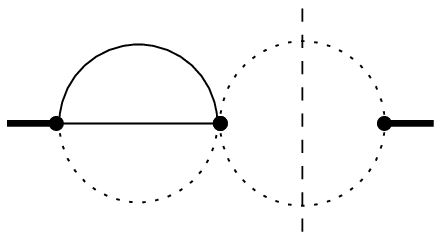} &&
&& \includegraphics[width=2cm,angle=0]{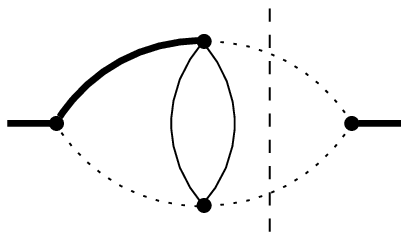} &&
&& \includegraphics[width=2cm,angle=0]{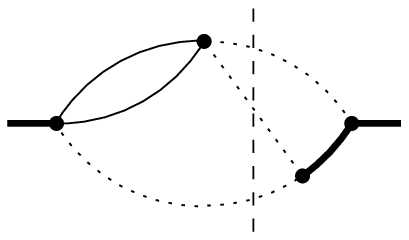} & \\[4mm]
$I_6$ &&&& $I_{12}$ &&&& $I_{18}$ &&& \\[-8mm]
&& \includegraphics[width=2cm,angle=0]{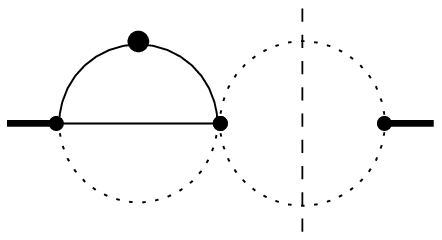} &&
&& \includegraphics[width=2cm,angle=0]{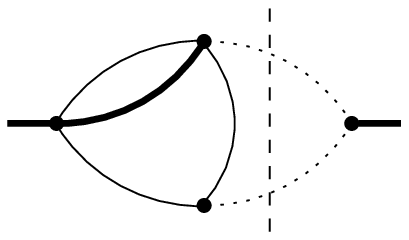} &&
&& \includegraphics[width=2cm,angle=0]{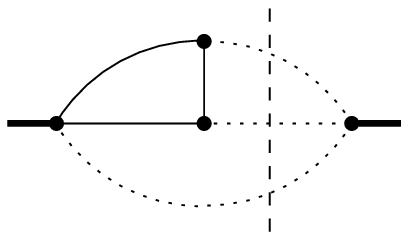} & \\\hline 
\end{tabular}
\caption{\sf Master integrals with unitarity cuts indicated by the dashed lines. 
Thick solid, thin solid and dotted internal lines denote propagators with masses
$m_b$, $m_c$ and zero, respectively. Thick solid external lines come with the momentum
$p$ such that $p^2=m_b^2$. Dots on the propagators indicate that they are squared. 
Arrows in $I_{16}$ define an irreducible numerator (see Eq.~(\ref{MIdef})).\label{tab:MI}}
\end{center}
\end{table}

\newsection{Master integrals and differential equations \label{sec:de}}

We generate the relevant Feynman diagrams using {\tt
FeynArts}~\cite{Hahn:2000kx}. The Dirac algebra after summing/averaging over
polarizations is performed with fully anticommuting $\gamma_5$, which is
allowed because only traces with even numbers of $\gamma_5$ can give
non-vanishing contributions in the considered problem. Once the results are
expressed in terms of scalar integrals, we reduce them to the MIs using
Integration By Parts (IBP) identities, with the help of the code {\tt
FIRE}~\cite{Smirnov:2014hma}. This way we express the six quantities on the
r.h.s. of Eqs.~(\ref{g27}) and (\ref{g27m}) as linear combinations of eighteen
master integrals $I_k$
\mathindent0cm 
\bea
\hat{G}_{27}^{(1)2P(d)}\!\!\!   &=&\!\! {\rm Re} \sum_{k=1 }^8    A_k I_k, \hspace{12.5mm} 
\hat{G}_{27}^{(1)2P(u)}          =      {\rm Re} \sum_{k=1 }^{13} B_k I_k, \hspace{12.5mm} 
\hat{G}_{27}^{(1)3P}             =      {\rm Re} \sum_{k=14}^{18} C_k I_k, \nnb\\[2mm]
\hat{G}_{27}^{(1)m,2P(d)}\!\!\! &=&\!\! {\rm Re} \sum_{k=1 }^9    D_k I_k, \hspace{9mm} 
\hat{G}_{27}^{(1)m,2P(u)}        =      {\rm Re} \sum_{k=1 }^{13} E_k I_k, \hspace{9mm} 
\hat{G}_{27}^{(1)m,3P}           =      {\rm Re} \sum_{k=14}^{18} F_k I_k. \label{sums}
\eea
\mathindent1cm 
with coefficients $A_k$--$F_k$ that are provided in the ancillary files~\cite{anc}.
The MIs to be determined are displayed in Tab.~\ref{tab:MI}.
In our normalization conventions for them, we include an extra factor of
$2^{11}\pi^5$ together with an appropriate integer power of $m_b$ that makes
each $I_k$ dimensionless. For instance,
\mathindent0cm
\bea
I_{13} &=& \mu^{6\ep}\!\! \int d{\rm PS}_2 \int 
\f{d^D q_1}{i(2\pi)^D} \f{d^D q_2}{i(2\pi)^D} \f{2^{11} \pi^5 m_b^2
}{[m_c^2 - (q_1+k_1)^2] [m_c^2 - q_1^2] [m_b^2 - q_2^2] [m_c^2 - (q_1+q_2-k_2)^2]^2},\nnb\\[3mm]
I_{16} &=& \f{2^{11} \pi^5}{m_b^4}\;
\mu^{6\ep}\!\! \int d{\rm PS}_3 \int \f{d^D q}{i(2\pi)^D} 
\f{-(pk_1)}{[m_c^2 - (q+k_2+k_3)^2][m_c^2 - q^2]},\label{MIdef}
\eea
\mathindent1cm
where $p$ is the external momentum, while $k_i$ are the cut propagator momenta
(directed from left to right in all the MIs of Tab.~\ref{tab:MI}). In the above
expressions, integrals over $d{\rm PS}_n$ stand for the $n$-body massless
phase space integrals, with $n=2,3$. They can be
simplified~\cite{GehrmannDeRidder:2003pne} to integrals over~ $\hat{s}_{ij} = 2
(k_i k_j)/m_b^2$~ for arbitrary integrands $W_2(\hat{s}_{12})$ and
$W_3(\hat{s}_{12}, \hat{s}_{13}, \hat{s}_{23})$:
\bea 
\mu^{2\ep}\!\! \int d{\rm PS}_2\; W_2 &=& 
\f{e^{\gamma\ep}\Gamma(1-\ep)}{8\pi\Gamma(2-2\ep)}
\int_0^1 \f{d\hat{s}_{12}}{\hat{s}_{12}^\ep}\; \delta(1-\hat{s}_{12})\, W_2 ~=~
\f{e^{\gamma\ep}\Gamma(1-\ep)}{8\pi\Gamma(2-2\ep)}\; W_2(1),\label{ph2}\\[3mm] 
\mu^{4\ep}\!\! \int d{\rm PS}_3\; W_3 &=& 
\f{m_b^2 e^{2\gamma\ep}}{2^7\pi^3\Gamma(2-2\ep)} 
\int_0^1\! \f{d\hat{s}_{12}}{\hat{s}_{12}^\ep}\! 
\int_0^1\! \f{d\hat{s}_{13}}{\hat{s}_{13}^\ep}\! 
\int_0^1\! \f{d\hat{s}_{23}}{\hat{s}_{23}^\ep}\;
\delta(1-\hat{s}_{12}-\hat{s}_{13}-\hat{s}_{23})\, W_3.
\eea
As already mentioned, we set $\mu^2 = e^\gamma m_b^2/(4\pi)$ when evaluating
the MIs. With the normalization as defined above, all the master integrals
$I_k$ and the coefficients $A_k,\ldots,F_k$ in Eq.~(\ref{sums}) are functions
of two variables only: $\ep$ and $z=m_c^2/m_b^2$. 

In four spacetime dimensions, the global normalization of all the considered
contributions to the decay rate is fixed by Eq.~(\ref{rate}). In $D=4-2\ep$
dimensions, we follow Ref.~\cite{Czakon:2015exa} where the global
normalization is fixed by
\be
\hat{G}_{77}^{(0)} ~=~ 8 \pi\, (1-\ep)\, \mu^{2\ep}\!\! \int d{\rm PS}_2 ~=~ 
\f{e^{\gamma\ep}\Gamma(2-\ep)}{\Gamma(2-2\ep)}.
\ee
The factor $(1-\ep)$ above comes from the Dirac algebra in
$\left| \me{s\gamma | Q_7 | b}\right|^2$ summed over polarizations.

To evaluate the MIs, we differentiate them with respect to $z$, and reduce the
resulting integrals to the MIs again, using the IBP identities. This gives us
a closed set of DEs which takes the following form ($I'_k = dI_k/dz$):
\bea 
&& \hspace{-8mm} 
I_1' = \fm{2-2\ep}{z} I_1,\hspace{1cm}
I_2' = \fm{1- \ep}{z} I_2,\hspace{1cm}
I_3' = \fm{1- \ep}{z} I_3,\hspace{1cm}
I_4' = \fm{2-2\ep}{z(1-4z)}(I_1-I_2) - \fm{2-4\ep}{1-4z} I_4,\nnb\\[2mm]
&& \hspace{-8mm} 
I_5' = -2 I_6,\hspace{2cm} 
I_6' =  \fm{(1-\ep)^2}{z^2 (1-4z)} I_1 -\fm{2 - 7\ep + 6\ep^2}{z(1-4z)} I_5  
       -\fm{(6-16\ep)z +\ep}{z(1-4z)} I_6,\nnb\\[2mm]
&& \hspace{-8mm} 
I_7' = -2 I_8,\hspace{2cm} 
I_8' =  \fm{(1-\ep)^2}{4z^2(1-z)} (I_1-2I_2) -\fm{(1-2\ep)(2-3\ep)}{4z(1-z)} I_7
       +\fm{1-4\ep-(3-8\ep)z}{2z(1-z)} I_8,\nnb\\[2mm]
&& \hspace{-8mm} 
I_9' = \fm{1-\ep}{z} (I_9-I_4) +\fm{2-3\ep}{z} I_7 + 2 I_8,\hspace{1cm} 
I_{10}' = -\fm{1}{z} I_6 + \fm{1-2\ep}{z} I_{10},\nnb\\[2mm]
&& \hspace{-8mm} 
I_{11}' = \fm{(1-\ep)^2}{(1-2\ep)z^2} I_1 + \fm{1-\ep}{z} I_4 + \fm{1-3\ep}{z} I_{11},\hspace{8mm} 
I_{12}' =  \fm{(1-\ep)^2}{2(1-2\ep)z^2} I_1 +\fm{1-\ep}{z} I_4 
          -\fm{2-3\ep}{2z} I_7 -I_8 -\fm{\ep}{z} I_{12} -I_{13},\nnb\\[2mm] 
&& \hspace{-8mm} 
I_{13}' =  \fm{(1-\ep)^2}{2z^2(1-z)(1-4z)} (3I_2 -\fm{(2-3z+4z^2)}{2z} I_1)
           +\fm{2-6\ep+4\ep^2}{z(1-4z)} I_4 -\fm{(1-2\ep)(2-3\ep)}{4z(1-z)} I_7 
           -\fm{(1-2\ep)(1 + z)}{2z(1-z)} I_8 -\fm{2\ep}{z} I_{13},\nnb\\[2mm] 
&& \hspace{-8mm} 
I_{14}' = \fm{1-\ep}{z} I_{14},\hspace{8mm}
I_{15}' = \fm{-2+2\ep}{z(1-4z)} I_{14} -\fm{2-2\ep+(2-4\ep)z}{z(1-4z)} I_{15} -\fm{6-8\ep}{z(1-4z)} I_{16},\hspace{8mm}
I_{16}' = \fm{1-\ep}{z} I_{15} + \fm{3-4\ep}{z} I_{16},\nnb\\[2mm] 
&& \hspace{-8mm} 
I_{17}' = -\fm{2-2\ep}{z(1-4z)} I_{14} -\fm{4-6\ep - 6(1-2\ep)z}{z(1-4z)} I_{15} 
          -\fm{6-8\ep}{z(1-4z)} I_{16} +\fm{1-3\ep}{z} I_{17},\nnb\\[2mm] 
&& \hspace{-8mm} 
I_{18}' = -\fm{1-\ep}{z^2(1-4z)}  I_{14}  -\fm{1-\ep +(1-2\ep)z}{z^2(1-4z)} I_{15}
          -\fm{3-4\ep}{z^2(1-4z)} I_{16}  -\fm{\ep}{z} I_{18}. \label{DEs}
\eea

The necessary boundary conditions for the above DEs are evaluated in the $z
\gg 1$ limit, in which the integrals become considerably simpler. Large-$z$
expansions of the three-body MIs ($I_{14}$--$I_{18}$) are obtained in a
straightforward manner using the Feynman parameterization. In the two-body case
($I_1$--$I_{13}$), we have used the code {\tt
exp}~\cite{Harlander:1997zb,Seidensticker:1999bb} that applies the
asymptotic expansion method in a fully automatic manner.  

\newsection{Numerical results for the counterterm contributions \label{sec:num}}

In our first approach, we expanded all the MIs in $\ep$ up to the relevant
orders, which gave us a system of coupled ordinary DEs for 76 functions of $z$
being a single variable. The boundary condition was chosen at $z=20$, and
evaluated using the expansions up to ${\mathcal O}(1/z^8)$. Since the
coefficients of the DEs~(\ref{DEs}) have singularities at $z \in
\{0,\f14,1\}$, the variable $z$ must be treated as complex, and a contour in
the complex $z$-plane must be chosen for the numerical solution. In practice,
we solved the DEs along ellipses in the {\em lower} half-plane of $z$,
following the Feynman prescription $m_c^2 \to m_c^2 - i\varepsilon$ in the
propagators, and treating $m_b^2$ in $z = m_c^2/m_b^2$ as a real rescaling
factor. In this way, non-vanishing imaginary parts in some of the MIs were
obtained on the real axis for $z < \f14$ (below the $c\bar c$ production
threshold), despite the fact that both the boundary conditions at $z=20$ and
the DE coefficients were purely real. In particular, the proper imaginary
parts of the functions $a(z)$ and $b(z)$ from Eqs.~(3.3)--(3.4) of
Ref.~\cite{Buras:2002tp} were recovered. Such imaginary parts drop out from
our final results in Eqs.~(\ref{g27})--(\ref{g27m}) but, nevertheless, they
provide convenient cross-checks of the calculation.

The numerical solution was obtained using the {\tt FORTRAN} code {\tt
ZVODE}~\cite{zvode}, upgraded to quadrupole-double precision with the help of
the {\tt QD} computation package~\cite{qd}.
%
%
The physical value of $z$ is around $0.06$. However, apart from the vicinity
of this point, we have considered a wide range of the final values, in order to
test the limits at $z \to 0$, as well as the behaviour around the threshold at
$z=\f14$.

\begin{figure}[t]
\begin{center}
\begin{tabular}{lcr}
\includegraphics[width=53mm,angle=0]{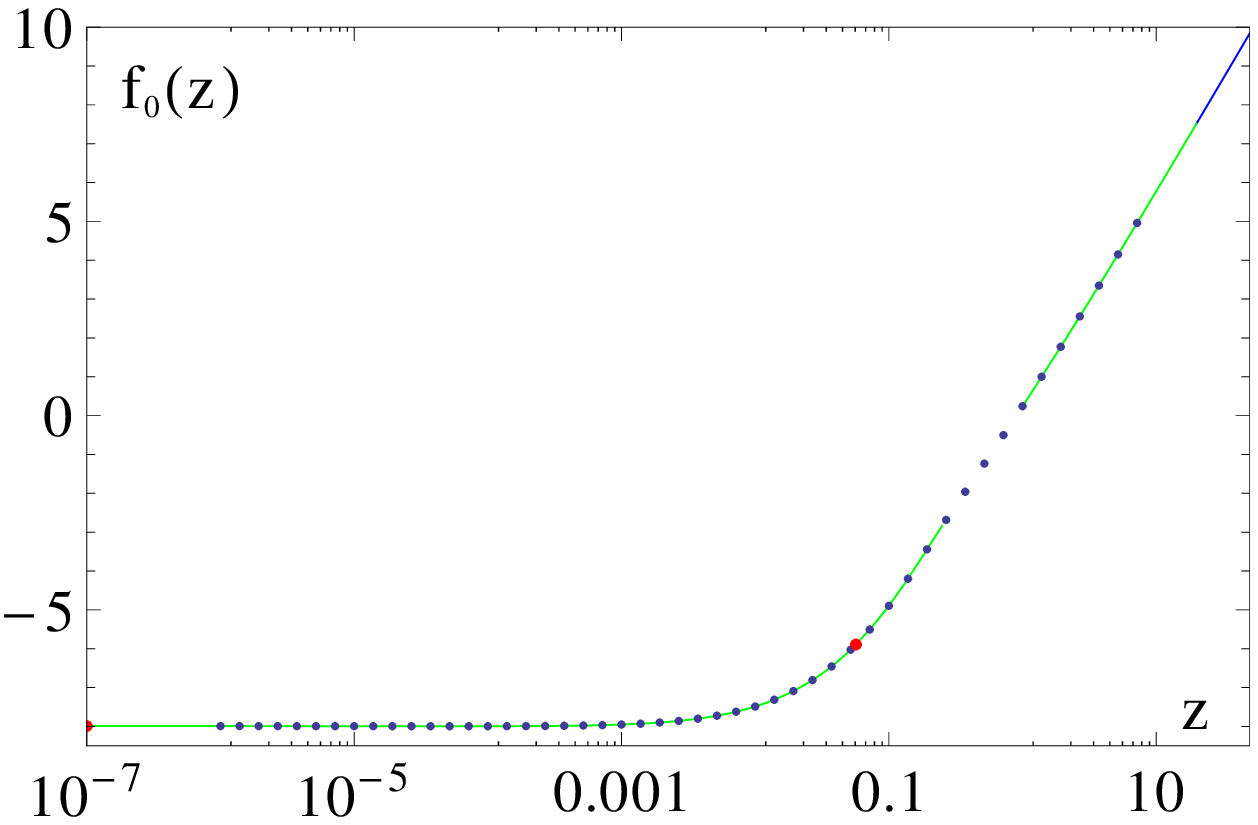}
\includegraphics[width=54mm,angle=0]{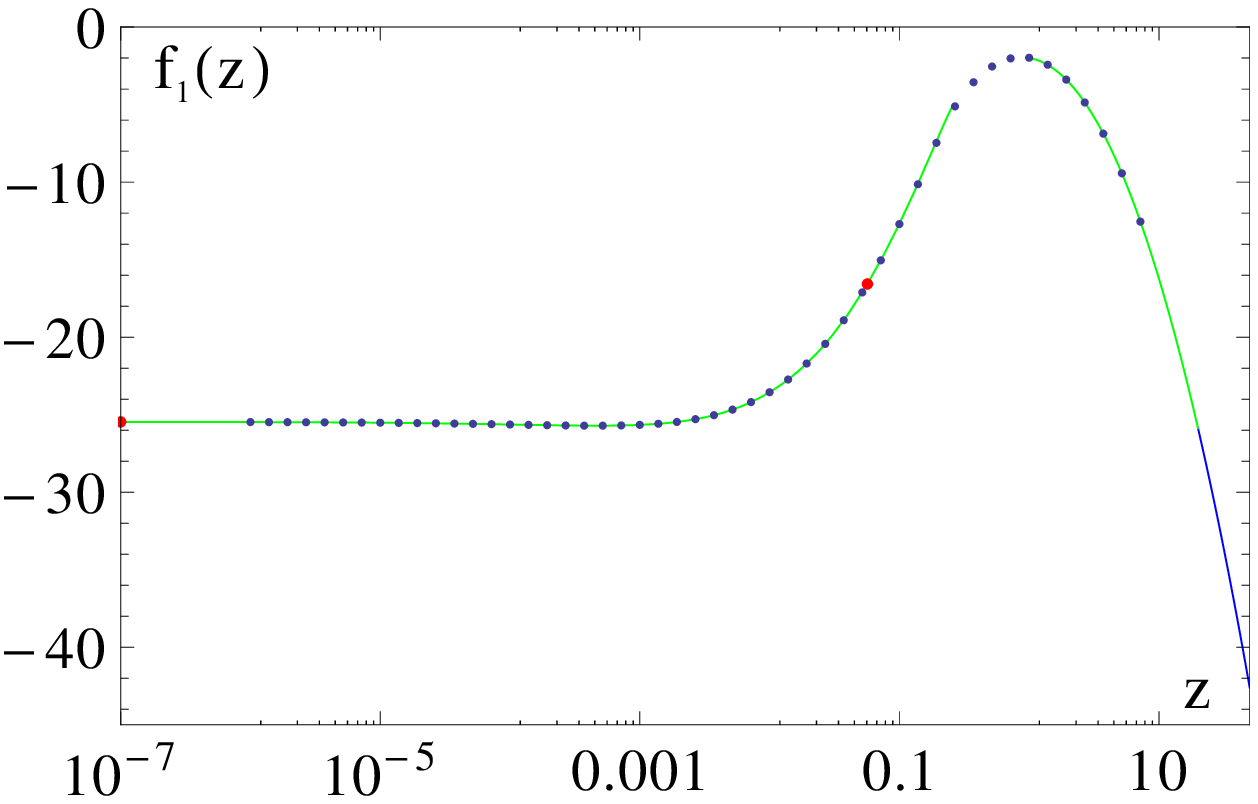}
\includegraphics[width=56mm,angle=0]{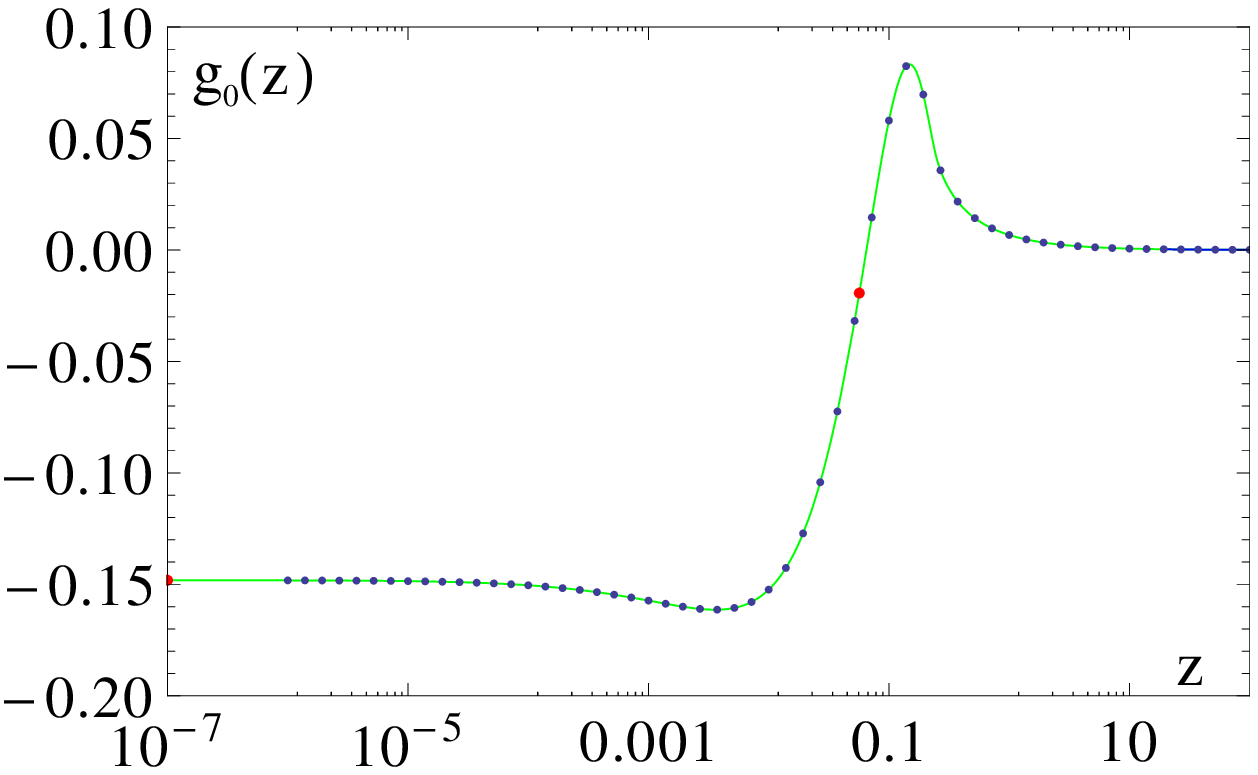}\\
\includegraphics[width=55mm,angle=0]{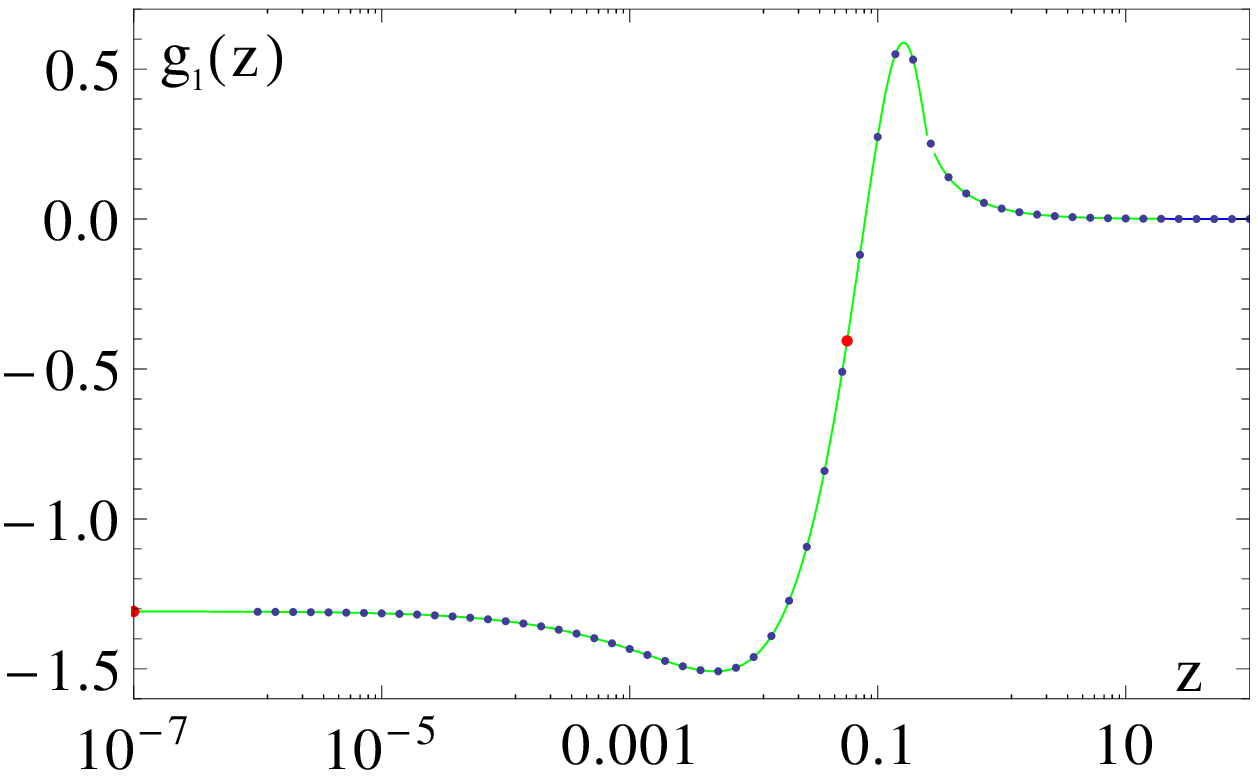}
\includegraphics[width=53mm,angle=0]{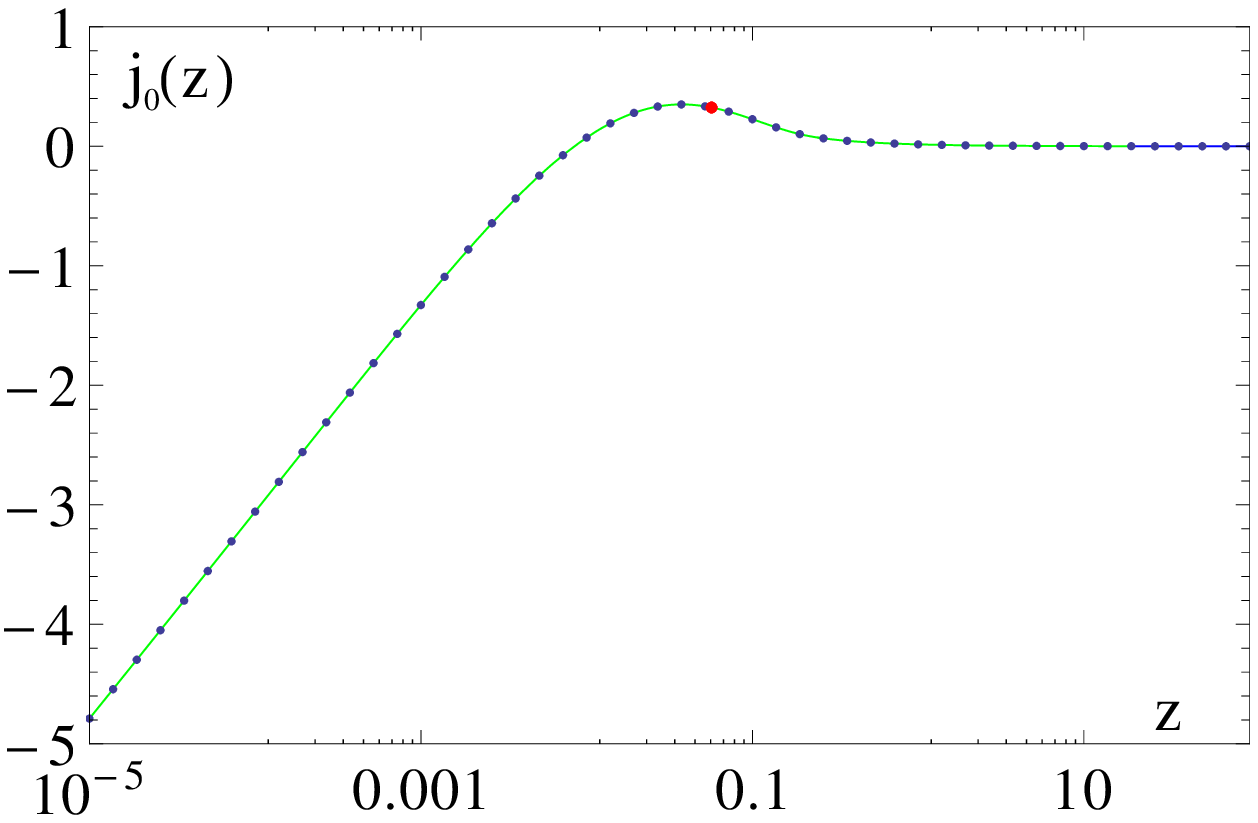}
\includegraphics[width=55mm,angle=0]{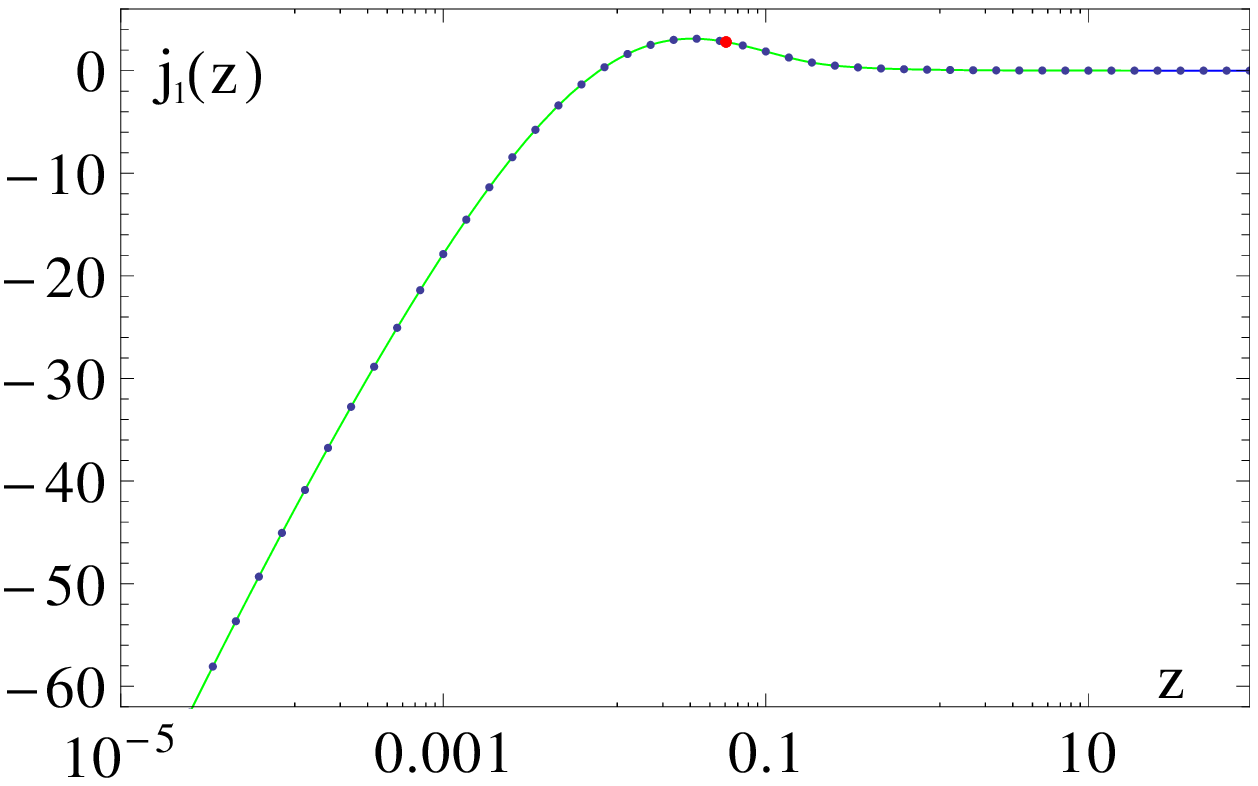}\\
\includegraphics[width=55mm,angle=0]{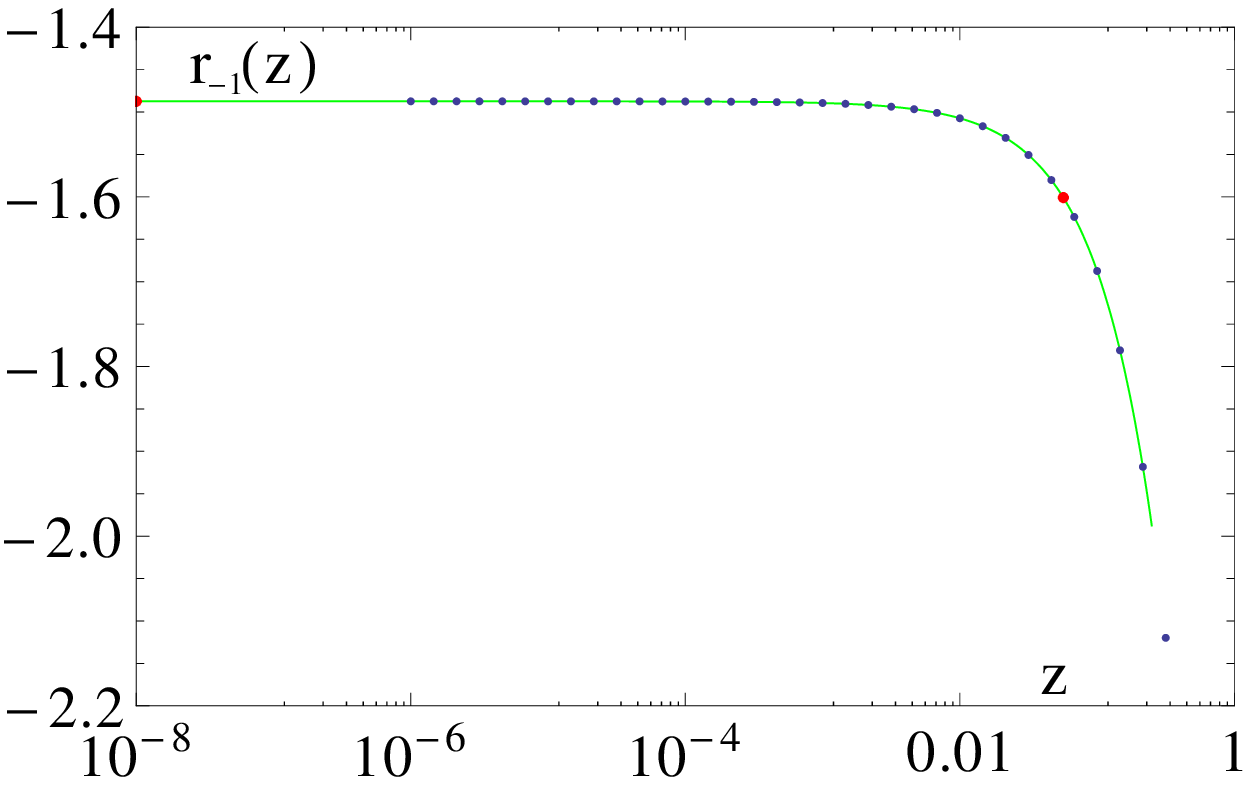}
\includegraphics[width=55mm,angle=0]{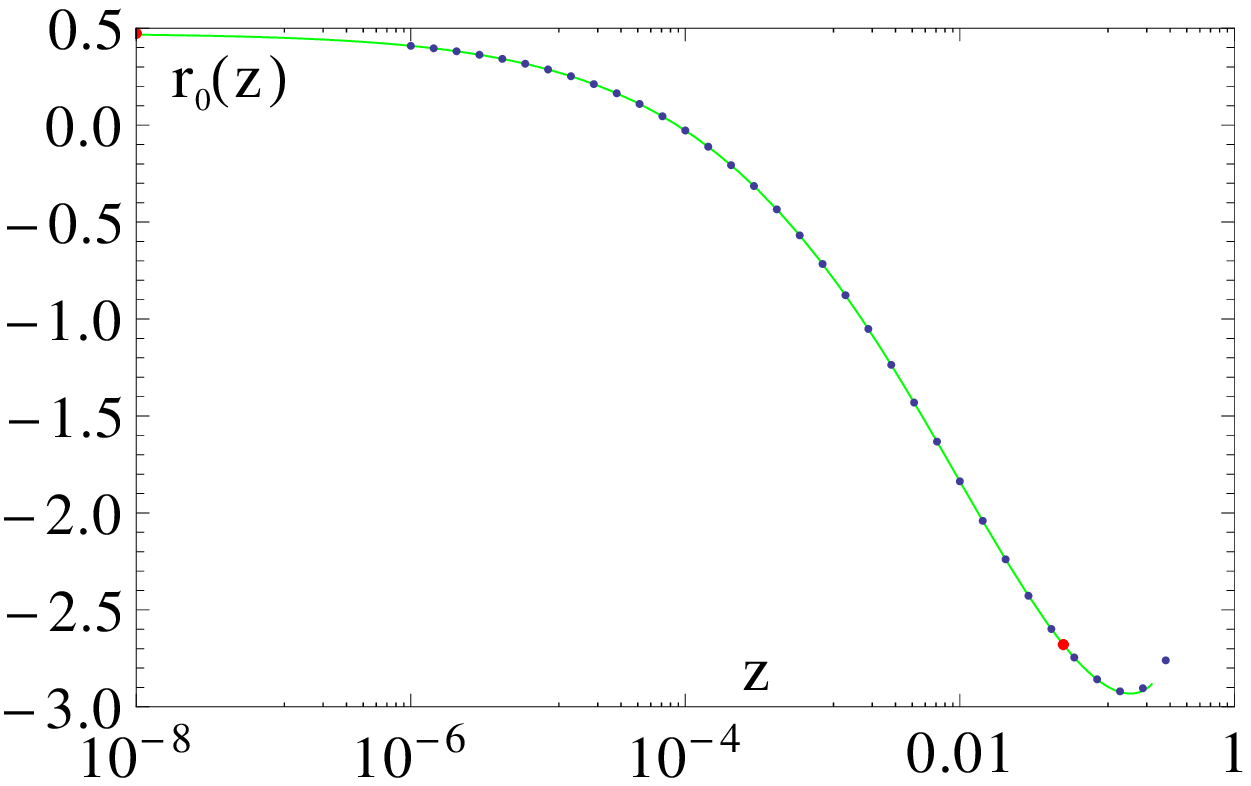}
\includegraphics[width=53mm,angle=0]{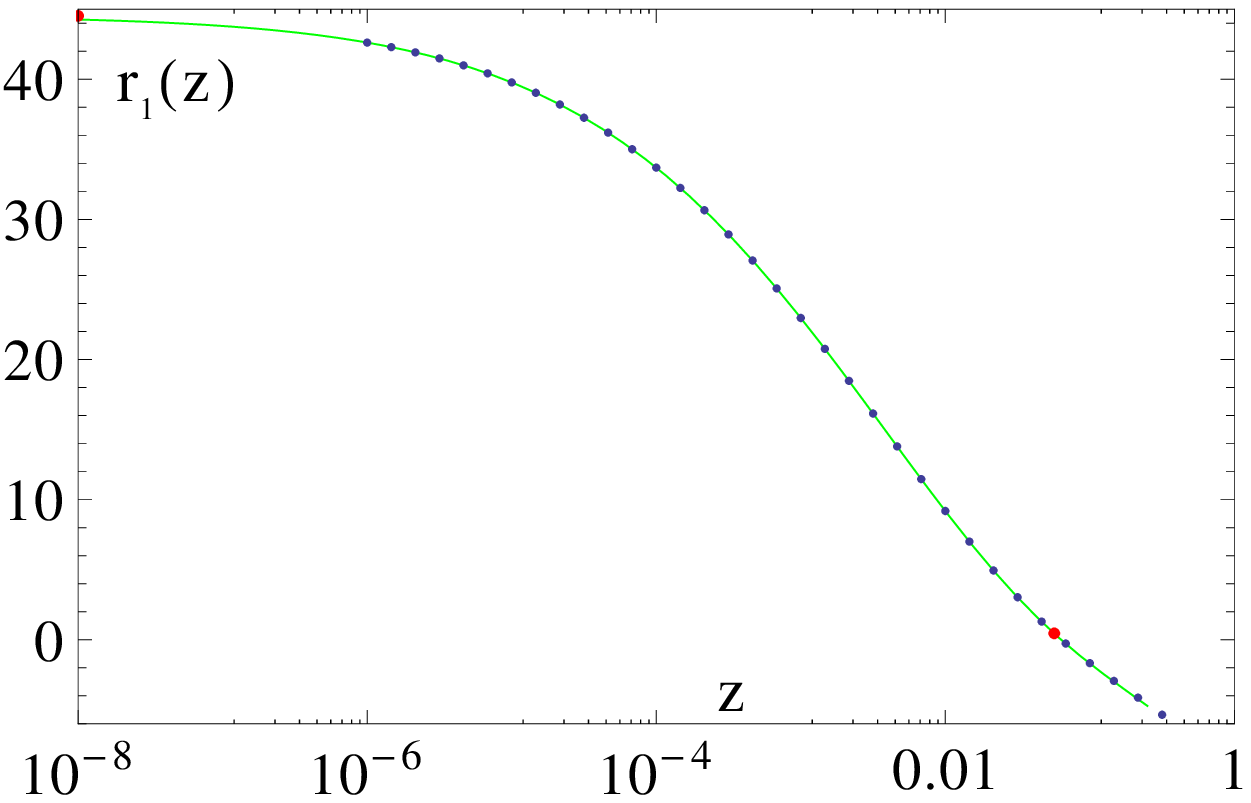}
\end{tabular}
\caption{\sf Plots of the functions defined in Eqs.~(\ref{def.f})--(\ref{def.r}) (see the text).\label{fig:z-plots}}
\end{center}
\end{figure}

To present our numerical results for the quantities listed in
Eq.~(\ref{sums}), we parameterize them in terms of nine functions of $z$ as
follows
\mathindent0cm
\bea
\hat{G}_{27}^{(1)2P} &\equiv& \hat{G}_{27}^{(1)2P(d)} + \hat{G}_{27}^{(1)2P(u)} ~=~ 
-\f{92}{81\ep} + f_0(z) + \epsilon f_1(z) + {\mathcal O}(\ep^2), \label{def.f}\\[2mm]
\hat{G}_{27}^{(1)3P} &=& g_0(z) + \epsilon g_1(z) + {\mathcal O}(\ep^2), \label{def.g}\\[2mm]
\hat{G}_{27}^{(1)m,3P} &=& j_0(z) + \epsilon j_1(z) + {\mathcal O}(\ep^2), \label{def.j}\\[2mm]
\hat{G}_{27}^{(1)m,2P} &\equiv& \hat{G}_{27}^{(1)m,2P(d)} + \hat{G}_{27}^{(1)m,2P(u)} ~=~
-\f{1}{3\ep^2} + \f{1}{\ep} r_{-1}(z) + r_0(z) + \epsilon r_1(z) + {\mathcal O}(\ep^2).\label{def.r}
\eea
Their dependence on $z$ is shown in Fig.~\ref{fig:z-plots}. Each of the (blue)
dots corresponds to a particular final value of the numerical solution of the
DEs along an ellipse in the complex plane. The physical point around $z=0.06$
is marked by a slightly bigger (red) dot. Similar (red) dots on the vertical
axes are used to indicate the limits at $z\to 0$ in all the cases when they
are finite, i.e. for all the functions except those in Eq.~(\ref{def.j}). They
can be taken over from the calculation performed at $z=0$ from the
outset~\cite{Czakon:2015exa}:
\bea
\left\{ f_0, f_1, g_0, g_1, r_{-1}, r_0, r_1 \right\} 
&\stackrel{\scriptscriptstyle z\to 0\;\;}{\longrightarrow}&
\left\{ -\fm{1942}{243},~~ 
        -\fm{26231}{729} + \fm{259}{243} \pi^2,~~
        -\fm{4}{27},~~
        -\fm{106}{81},~~ 
        -1-\fm{4}{81} \pi^2, 
\right. \nnb\\[2mm] && \hspace{2mm} \left.
        \fm{35}{9} -\fm{161}{972}\pi^2 -\fm{40}{27} \zeta_3,~~~~
        \fm{2521}{54} + \fm{2135}{2916} \pi^2 - \fm{65}{81} \zeta_3 - \fm{7}{81} \pi^4 \right\}. 
\eea
\mathindent1cm
where $\zeta_k\equiv \zeta(k)$. In the case of $\hat{G}_{27}^{(1)m,3P}$, one finds~\cite{Czakon:2015exa} 
\be
\hat{G}_{27}^{(1)m,3P}(z=0) = \fm{20}{27\ep} + \fm{770}{81} + 
\left( \fm{18191}{243} - \fm{35}{27} \pi^2 \right) \ep + {\mathcal O}(\ep^2).
\ee
The above expression contains a $1/\ep$ divergence, contrary to
Eq.~(\ref{def.j}). It arises due to the logarithmic divergences of the
functions $j_i(z)$ at $z\to 0$. Such non-commutation of limits is frequently
encountered in the framework of dimensional regularization.

Our expansions above $z=20$ that have served as the boundary conditions for
the DEs are displayed as solid (blue) lines in the first six plots of
Fig.~\ref{fig:z-plots}. The remaining (green) solid lines describe expansions
either in $z$ (for $z < \f14$) or in $\f{1}{z}$ (for $z > \f14$) that have
been obtained in our second approach -- see
Section~\ref{sec:fin}.  For $f_i(z)$, the expansion plots are
terminated away from the threshold at $z=\f14$, as they become inaccurate in
its vicinity. In the three-body cases ($g_i(z)$, $j_i(z)$) the applied
expansions are so accurate that their mismatch at $z=\f14$ is invisible within the
plot resolution.\footnote{
The solid lines for $g_0(z)$ actually correspond to the fully analytical result of Eq.~(\ref{g0}).}

We refrain from presenting here any numerical fits for the functions plotted
in Fig.~\ref{fig:z-plots}.  Instead, either expansions or exact expressions
for them will be given in Section~\ref{sec:fin}. It is evident from the plots
that our expansions around $z=0$ are sufficiently precise for phenomenological
purposes in the vicinity of the physical point.

\newsection{\boldmath Expansions in $z$ \label{sec:fin}}

In this section, we collect our final results for the small-$z$ expansions of
the functions defined in Eqs.~(\ref{def.f})--(\ref{def.r}).  They have been
obtained in Ref.~\cite{ARthesis} with the help of Feynman parameterizations,
Mellin-Barnes techniques~\cite{Boos:1990rg}, as well as our differential
equations in Eq.~(\ref{DEs}) that allowed to extend the expansions to higher
orders.

For $f_0(z)$, we properly recover the NLO functions $a(z)$ and $b(z)$ known
from the previous
calculations~\cite{Greub:1996jd,Greub:1996tg,Buras:2001mq,Buras:2002tp},
namely
\be
f_0(z) = -\f{1942}{243} + 2 {\rm Re}[ a(z) + b(z)].
\ee
The expansions of $a(z)$ and $b(z)$ up to ${\mathcal O}(z^6)$ are given in
Eqs.~(3.8)--(3.9) of Ref.~\cite{Buras:2002tp}, and we shall not repeat them
here. For $f_1(z)$, we obtain
\mathindent0cm
\bea
f_1(z) &=& -\fm{26231}{729} + \fm{259}{243} \pi^2 
+ \left[ 104 -\fm{472}{27}\pi^2 - \fm{496}{9} \zeta_3 -\fm{92}{405} \pi^4
+ \left( \fm{448}{9} -\fm{296}{27} \pi^2 - 16 \zeta_3\right) L
\right. \nnb\\[2mm] &-& \left.
\left( \fm{8}{3} -\fm{28}{27} \pi^2 \right) L^2 + \fm{8}{27} L^3 -\fm{10}{27} L^4 \right] z
+ \left[ \fm{1664}{81} -\fm{256}{27} \ln 2 -\fm{64}{27} L \right] \pi^2 z^{3/2}
-\fm{1120}{81} \pi^2 z^{5/2}\nnb\\[2mm] 
&+& \left[ \fm{296}{9} + \fm{40}{27} \pi^2 + \fm{32}{3} \zeta_3 + \fm{80}{81} \pi^4
+ \left( \fm{16}{9} -\fm{128}{27} \pi^2 \right) L - \left( \fm{16}{9} - \fm{16}{27} \pi^2 \right) L^2 
+ \fm{40}{27} L^3 -\fm{10}{27} L^4 \right] z^2\nnb\\[2mm] 
&+& \left[ \fm{22381}{729} -\fm{2180}{243} \pi^2 + \fm{304}{27} \zeta_3
- \left( \fm{1382}{243} - \fm{128}{81} \pi^2 \right) L 
- \fm{260}{27} L^2 + \fm{56}{9} L^3 \right] z^3 + {\mathcal O}(z^4),
\eea
where $L=\ln z$. The above result is in perfect agreement with Eqs.~(15)--(17)
of Ref.~\cite{Asatrian:2005pm}. As far as $g_0(z)$ is concerned, a fully
analytical expression is available
\be \label{g0}
g_0(z) = \left\{ \!\!\begin{array}{ll}
-\f{4}{27} - \f{14}{9} z + \f83 z^2 + \f83 z (1-2z)\, s\, X\, + \f{16}{9} z (6z^2-4z+1)\left(\f{\pi^2}{4}-X^2\right),
& \mbox{for~} z \leq \f14, \\[2mm]
-\f{4}{27} - \f{14}{9} z + \f83 z^2 +\f83 z (1-2z)\, t\, Y\, +\f{16}{9} z (6z^2-4z+1)\, Y^2,
& \mbox{for~} z > \f14, \end{array} \right.
\ee
where~ $s = \sqrt{1-4z}$,~ $X = \ln\f{1+s}{2} - \f12 L$,~ $t = \sqrt{4z-1}$,~ and~ $Y=\arctan(1/t)$.
For $g_1(z)$, we find
\bea
g_1(z)\!\! &=&\!\! -\fm{106}{81} 
- \left[ \fm{194}{9} -\fm{38}{9}\pi^2 -\fm{40}{9} \zeta_3
+ \left( 12 -\fm{4}{9} \pi^2 \right) L + \fm{20}{9} L^2 - \fm{8}{27} L^3 \right] z
+ \left[ \fm{124}{9} -8 \pi^2 -\fm{32}{9} \zeta_3 +\fm{200}{9} L
\right. \nnb\\[2mm] &+& \left.
\fm{16}{9} \pi^2 L  +\fm{8}{3} L^2 -\fm{64}{27} L^3 \right] z^2 
+ \left[ \fm{388}{81} -\fm{4}{3} \pi^2 +16 \zeta_3 -\fm{232}{27} L -\fm{4}{3} L^2  +\fm{8}{3} L^3 \right] z^3 
+ {\mathcal O}(z^4). \label{g1} 
\eea
It cannot be derived from Eq.~(23) in Ref.~\cite{Asatrian:2005pm} that has
been obtained with the help of phase-space integrations in four dimensions (as
stated above Eq.~(19) there). Our result corresponds to $D=4-2\ep$ throughout
the calculation, as required by the so-called reverse unitarity
method~\cite{Anastasiou:2002yz} that we have applied for the IBP in the
three-body case.

The remaining five functions correspond to quantities that have not been considered
in Ref.~\cite{Asatrian:2005pm} because they are unrelated to the charm quark mass
renormalization. Our results for their expansions in $z$ read
\bea
j_0(z) &=& \fm{302}{81} +\fm{20}{27} L 
+\left[ \fm{7}{3} -\fm{10}{9} \pi^2 +\fm{16}{9} \zeta_3 +\fm{2}{135} \pi^4
-\left( \fm{2}{3} -\fm{4}{9} \pi^2 +\fm{32}{9} \zeta_3 \right) L
+\left( \fm{10}{9} -\fm{4}{9} \pi^2 \right) L^2
-\fm{4}{27} L^3 
\right.\nnb\\[2mm] &+& \left.
\fm{2}{27} L^4 \right] z
+\left[ -\fm{44}{9} +\fm{8}{9} \pi^2 +\fm{8}{3} L -\fm{8}{9} L^2 \right] z^2
+\left[ -\fm{32}{81} -\fm{4}{9} \pi^2 -\fm{16}{27} L +\fm{4}{9} L^2 \right] z^3
+ {\mathcal O}(z^4),\nnb\\[3mm]
j_1(z) &=& \fm{11225}{243} -\fm{20}{27} \pi^2 +\fm{52}{9} L -\fm{10}{27} L^2 
+\left[ 2 -\fm{35}{9} \pi^2 +12 \zeta_3 -\fm{4}{45} \pi^4 + 16 \zeta_5 + \fm{8}{9} \pi^2 \zeta_3 -\left( \fm{208}{9} 
\right.\right.\nnb\\[2mm] &-& \left.\left.
\fm{14}{3} \pi^2 + 24 \zeta_3 -\fm{46}{135} \pi^4 \right) L
+\left( \fm{44}{9} -\fm{10}{3} \pi^2 +\fm{40}{9} \zeta_3 \right) L^2
-\left( \fm{8}{3} -\fm{16}{27} \pi^2 \right) L^3
+\fm{2}{3} L^4 -\fm{14}{135} L^5 \right] z\nnb\\[2mm] 
&+& \left[ -\fm{284}{9} -\fm{4}{3} \pi^2 +\fm{112}{3} \zeta_3
        +\left( \fm{8}{3} + 8 \pi^2 \right) L -\fm{8}{3} L^2 -\fm{16}{9} L^3 \right] z^2\nnb\\[2mm] 
&+& \left[ \fm{4264}{243} -\fm{110}{27}\pi^2 -\fm{8}{3}\zeta_3 
        -\fm{1528}{81} L +\fm{134}{27} L^2 -\fm{4}{9} L^3 \right] z^3
+ {\mathcal O}(z^4),\nnb\\[3mm]
r_{-1}(z) &=& -1 - \fm{4}{81} \pi^2 - 2z,\nnb\\[3mm]
r_0(z) &=& \fm{35}{9} - \fm{161}{972} \pi^2 -\fm{40}{27} \zeta_3
-\fm{64}{9} \pi^2 z^{1/2} +\fm{208}{81} \pi^2 z^{3/2}\nnb\\[2mm]
&+&\left[ \fm{152}{9} +\fm{32}{27} \pi^2 -\fm{64}{9} \zeta_3 +\fm{8}{135} \pi^4
-\left( \fm{56}{3} + \fm{80}{27} \pi^2 \right) L 
+\left( \fm{8}{3} +\fm{4}{27}  \pi^2 \right) L^2
-\fm{32}{27} L^3 +\fm{1}{27} L^4 \right] z\nnb\\[2mm]
&+&\left[ -\fm{302}{81} -\fm{62}{27} L +\fm{64}{9} L^2 +\fm{26}{9} \pi^2 \right] z^2
+\left[ -\fm{968803}{36450} +\fm{26759}{1215}  L +\fm{952}{81}  L^2 +\fm{952}{243} \pi^2 \right] z^3
+ {\mathcal O}(z^4),\nnb\\[3mm]
r_1(z) &=& \fm{2521}{54} +\fm{2135}{2916} \pi^2 -\fm{65}{81} \zeta_3 -\fm{7}{81} \pi^4
-\left[ \fm{448}{9}-\fm{512}{9} \ln 2 -\fm{128}{9} L \right] \pi^2 z^{1/2}
+\left[ C -\left( \fm{184}{3} +\fm{88}{9} \pi^2 
\right.\right.\nnb\\[2mm] &+& \left.\left.
\fm{608}{9} \zeta_3 -\fm{58}{405} \pi^4 \right) L
+\left( \fm{260}{9} +\fm{68}{27} \pi^2 +\fm{32}{9} \zeta_3 \right) L^2
-\left( \fm{224}{27} +\fm{16}{81} \pi^2 \right) L^3 +\fm{47}{27} L^4 -\fm{8}{135} L^5 \right] z\nnb\\[2mm]
&+&\left[ \fm{3904}{81} - \fm{1664}{81} \ln 2 -\fm{416}{81} L \right] \pi^2 z^{3/2}
-\left[ \fm{4535}{243} -\fm{74}{27}\pi^2 - 12\zeta_3
       + \left( \fm{1273}{81} +\fm{8}{9} \pi^2 \right) L -\fm{163}{9} L^2 
\right.\nnb\\[2mm] &+& \left.
8 L^3 \right] z^2 -\fm{10672}{6075}  \pi^2 z^{5/2}
-\left[ \fm{14697769}{273375} -\fm{45847}{3645}\pi^2 +\fm{1904}{81} \zeta_3
       -\fm{1402729}{36450} L -\fm{1082}{243} L^2 +\fm{952}{81} L^3 \right] z^3\nnb\\[2mm]
&+& {\mathcal O}(z^{7/2}).
\label{r1} \eea
\mathindent1cm
In the latter equation, the constant $C \simeq -292.228$ has been determined only numerically.

Eqs.~(\ref{g1})--(\ref{r1}) do not explicitly show all the terms
that have been taken into account for the plots in Fig.~\ref{fig:z-plots}. 
In the latter case, we used the series terminating at 
    $z^{11}$, $z^{10}$, $z^{10}$, $z^5$  and $z^5$
for $g_1$,    $j_0$,    $j_1$,    $r_0$  and $r_1$, respectively.
Extended small-$z$ expansions, large-$z$ expansions, as well as fully analytical
results for many of our master integrals can be found in Ref.~\cite{ARthesis}.

It is likely that fully analytical results for all the considered counterterm
contributions could be found using the so-called canonical
basis~\cite{Henn:2013pwa,Lee:2014ioa} of the MIs, and solving the DEs
order-by-order in $\ep$, in terms of iterated integrals. However, at the time
the main part of our calculation was performed, no automatic tools for
constructing canonical bases at three loops (see, e.g.,
\cite{Gituliar:2017vzm,Prausa:2017ltv}) were available. On the other hand,
the expansions in $z$ are certainly sufficient for all the phenomenological
purposes in our case, while the numerical method is most suitable for
preparing the bare calculation.

\newsection{Summary \label{sec:sum}}

We calculated all the relevant counterterm contributions to
$\hat{G}^{(2)}_{17}$ and $\hat{G}^{(2)}_{27}$ at the NNLO for the physical
value of the charm quark mass $m_c$. In combination with the future bare
calculation of these quantities, the current results are necessary for removing
one of the most important perturbative uncertainties in the SM prediction
${\mathcal B}_{s\gamma}$. In our first approach, we derived differential
equations for the master integrals, and solved them numerically using the
boundary conditions at large $m_c$. In the second approach, a combination of
various techniques was used to determine the master integrals in terms of
power-logarithmic expansions around $m_c=0$. In most of the cases, logarithmic
divergences at $m_c \to 0$ are absent, and our results properly converge to
the $m_c=0$ ones previously found in Ref.~\cite{Czakon:2015exa}.

\section*{Acknowledgments}

We are grateful to Micha{\l} Czakon and Paul Fiedler for their advice
concerning numerical solutions of the differential equations, and for sharing
their codes with us. M.M. and A.R. acknowledge support from the National
Science Centre (Poland) research project, decision no
DEC-2014/13/B/ST2/03969. The research of M.S. has been supported by the BMBF
grant 05H15VKCCA.

\end{document}